\documentclass[%
 reprint,
 amsmath,amssymb,
 aps,
]{revtex4-1}

\usepackage{graphicx}
\usepackage{dcolumn}
\usepackage{bm}
\usepackage[mathlines, columnwise]{lineno}%
\usepackage{hyperref}

\usepackage{braket}

\newcounter{lastnote}

\begin{document}

\preprint{APS/123-QED}

\title{Monitoring and manipulating Higgs and Goldstone modes in a supersolid quantum gas}

\author{Julian L\'eonard}
\author{Andrea Morales}
\author{Philip Zupancic}
\author{Tobias Donner}
\email{donner@phys.ethz.ch}
\author{Tilman Esslinger}
\affiliation{Institute for Quantum Electronics, ETH Z\"urich, CH-8093 Z\"urich, Switzerland}


\begin{abstract}
Access to collective excitations lies at the heart of our understanding of quantum many-body systems. We study the Higgs and Goldstone modes in a supersolid quantum gas that is created by coupling a Bose-Einstein condensate symmetrically to two optical cavities. The cavity fields form a U(1)-symmetric order parameter that can be modulated and monitored along both quadratures in real time. This enables us to measure the excitation energies across the superfluid-supersolid phase transition, establish their amplitude and phase nature, as well as characterize their dynamics from an impulse response. Furthermore, we can give a tunable mass to the Goldstone mode at the crossover between continuous and discrete symmetry by changing the coupling of the quantum gas with either cavity. 
\end{abstract}

\maketitle

Collective excitations are crucial for describing the dynamics of quantum many-body systems. They provide unified explanations of phenomena studied in different disciplines of physics, such as in condensed matter \cite{Pekker2015} or particle physics \cite{Bernstein1974}, or in cosmology \cite{Lemoine2008}. The symmetry of the underlying effective Hamiltonian determines the character of the excitations, which changes in a fundamental way when a continuous symmetry is broken at a phase transition. Excitations can now appear both at finite and zero energy. 

\begin{figure}[h!]
	\centering
		\includegraphics[width=\columnwidth]{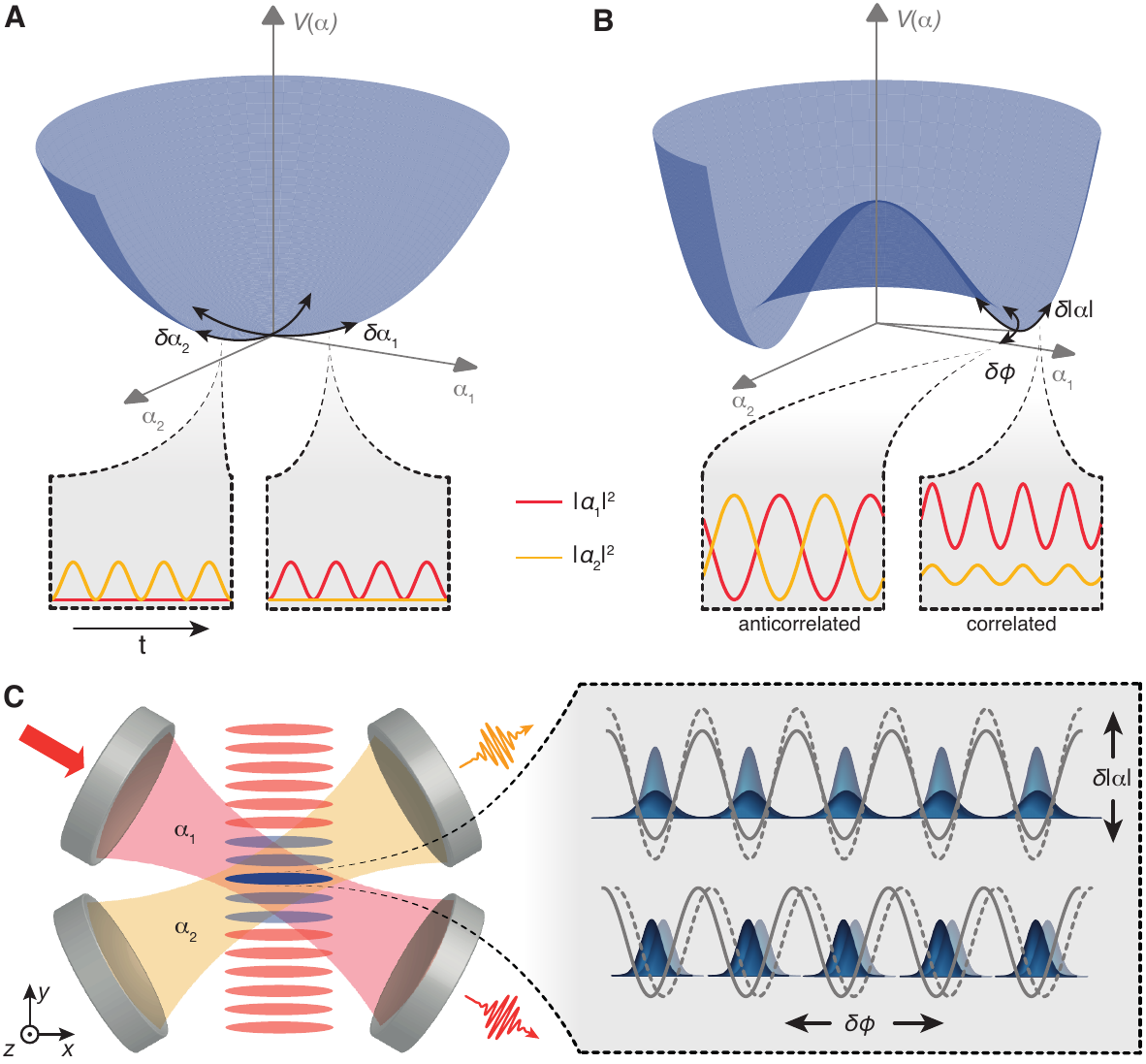}
	\caption{ \textbf{Higgs and Goldstone modes for a U(1) symmetry.} Effective potential across the phase transition as a function of the order parameter $\alpha=\alpha_1 + i \alpha_2=|\alpha| e^{i \phi}$. \textbf{(A)} In the normal phase, two excitations, $\delta\alpha_1$ and $\delta\alpha_2$, correspond to fluctuations of the order parameter along each quadrature. Both excitations have an amplitude character involving one quadrature of the order parameter. \textbf{(B)}. In the ordered phase, Higgs and Goldstone modes describe amplitude ($\delta|\alpha|$) and phase ($\delta\phi$) fluctuations around a finite expectation value of the order parameter. The squares of the quadratures show either correlations (Higgs) or anticorrelations (Goldstone). \textbf{(C)} Illustration of the experiment. A Bose-Einstein condensate (blue stripes) cut into slices by a transverse pump lattice potential (red stripes) enters a supersolid state and breaks translational symmetry along $x$ by symmetrically coupling it to two optical cavity modes $\alpha_1$ (red) and $\alpha_2$ (yellow) with a transverse pump lattice along $y$. The emerging Higgs and Goldstone excitations correspond to fluctuations of the strength and position of the density modulation, as shown in the zoom-in for one slice. They can be excited and read out with probe pulses on each cavity.}
	\label{fig:scheme}
\end{figure}

In the paradigmatic case of models with U(1)-symmetry breaking, the system can be described by a complex scalar order parameter in an effective potential as illustrated in Fig.~\ref{fig:scheme}(A-B) \cite{Sachdev2011}. In the normal phase, the potential is bowl-shaped with a single minimum at vanishing order parameter, and correspondingly two orthogonal amplitude excitations. Within the ordered phase, the potential shape changes to a 'sombrero' with an infinite number of minima on a circle. Here, fluctuations of the order parameter reveal two different excitations: a Higgs (or amplitude) mode, which stems from amplitude fluctuations of the order parameter and shows a finite excitation energy, and a Goldstone (or phase) mode, which stems from phase fluctuations of the order parameter and has zero excitation energy. The former should yield correlated fluctuations in the two squared quadratures of the order parameter, whereas the latter should show anticorrelated behavior. 

\begin{figure*}[t]
	\centering
		\includegraphics[width=120mm]{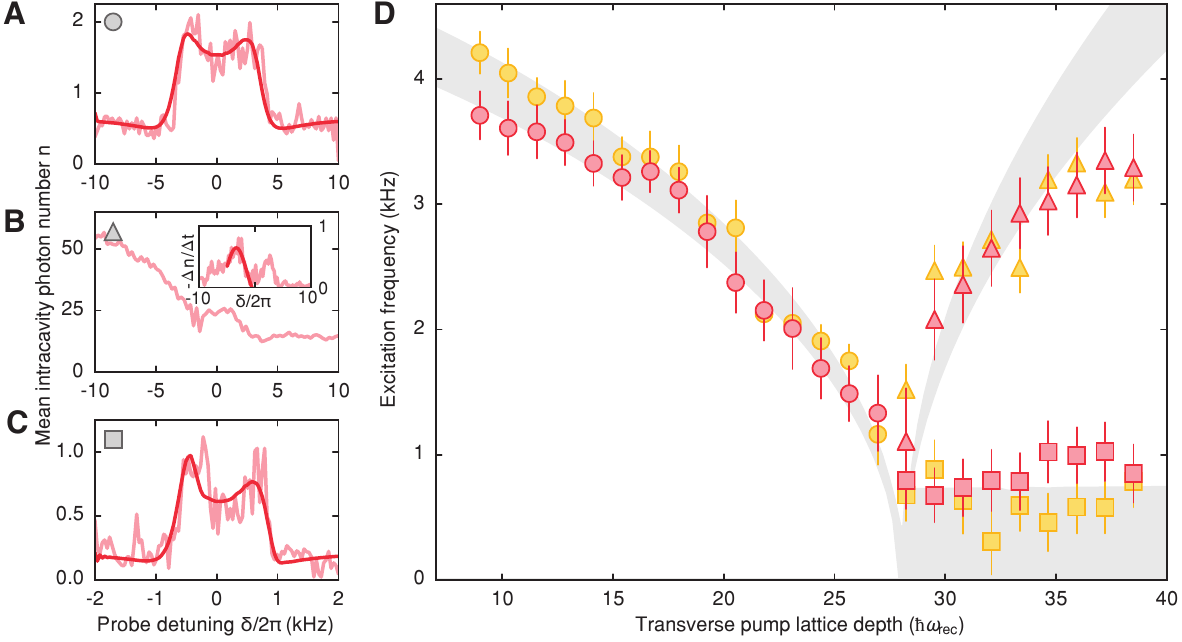}
	\caption{\textbf{Excitation spectrum across the phase transition. (A-C)} Response of the intracavity photon number to the probe field. Shaded red lines show the mean photon numbers for cavity 1, binned in intervalls of $0.2\,\text{ms}$ and averaged over at least ten realizations. Solid lines show fits from a theoretical model \cite{SI_excitations}. \textbf{(A)} Response to a probe field of $\bar{n}_1=3.4(2)$ mean photons in cavity 1 whose frequency is ramped by $1\,\text{kHz}/\text{ms}$, measured at $16.7(4)\,\hbar\omega_\text{rec}$ lattice depth. \textbf{(B)} Photon number during a probe field of $\bar{n}_2=3.4(1)$ in cavity 2 whose frequency is ramped by $1\,\text{kHz}/\text{ms}$, measured at $35.9(8)\,\hbar\omega_\text{rec}$ lattice depth. The inset displays the inferred negative derivative, representing the response as a function of $\delta$ with symmetric resonances at positive and negative detunings. The fit takes into account the negative resonance only to limit influence from the decaying order parameter. \textbf{(C)} Response to a probe field of $\bar{n}_1=0.06(1)$ in cavity 1 whose frequency is ramped by $0.2\,\text{kHz}/\text{ms}$, measured at $35.9(8)\,\hbar\omega_\text{rec}$ lattice depth. \textbf{(D)} Resonance frequencies for the normal phase (circles) and the ordered phase at high (triangles) and low (squares) frequencies extracted from the response to probe pulses in cavity 1 (red) and 2 (yellow). The gray-shaded area shows the theoretical prediction including experimental uncertainties \cite{SI_excitations}. Errorbars combine fit errors and the uncertainty of the probe frequency.}
	\label{fig:spectrum}
\end{figure*}

Condensed matter systems typically do not provide access to both quadratures of the order parameter, and Higgs and Goldstone modes have to be excited and detected by incoherent processes. In addition, the idealized situation is often disguised by further interactions that reduce the number of distinct modes \cite{Podolsky2011, Pekker2015}. For charged particles, the minimal coupling to a vector potential can even completely suppress the Goldstone mode through the Anderson-Higgs mechanism \cite{Bernstein1974}. In charge-density wave compounds, a persisting Higgs mode has been observed as a well-defined resonance \cite{Sooryakumar1980, Littlewood1981, Measson2014}. In superfluid Helium \cite{Yarnell1958} and Bose-Einstein condensates \cite{Stamper-Kurn1999} the coupling between amplitude and phase excitations spares the Goldstone mode only \cite{Varma2002}. For cases of only weak coupling between the modes, a Higgs mode has been identified as a broad resonance in antiferromagnets \cite{Ruegg2008}, and as a resonance-free onset of the response to lattice modulation in two-dimensional optical lattices \cite{Endres2012, Liu2015}. 

We report on the observation and manipulation of a Higgs and a Goldstone mode and directly verify their amplitude and phase character by monitoring the dynamics of each quadrature of the order parameter in real-time. The experimental situation is illustrated in Fig.~\ref{fig:scheme}C. A spatial U(1)-symmetry breaking is induced in a setting in which a Bose-Einstein condensate is off-resonantly driven by a transverse optical standing wave and degenerately coupled to the modes of two optical cavities \cite{Leonard2017}. Both cavity modes overlap with the Bose-Einstein condensate and are oriented in a $60^\circ$ angle with respect to the transverse pump lattice, which also provides an attractive one-dimensional standing wave potential for the atoms. Coherent transitions, induced by a transverse pump photon plus a cavity photon, couple two states with and without a density-modulation of the condensate \cite{SI_excitations}. For small two-photon couplings the system remains in the normal phase. As soon as the coupling strength exceeds a critical value, the kinetic energy associated with the density-modulation of the atomic wave function is overcome and the system enters a self-organized phase, which is periodically ordered perpendicular to the transverse pump lattice. Due to the symmetric coupling to both cavities, this phase transition breaks the continuous translational symmetry along the $x$-axis and a supersolid phase emerges \cite{Leonard2017}.

Since the density ordering is driven by light scattering between the transverse pump lattice and the cavities, the phase transition is accompanied by the appearance of non-zero real-valued field amplitudes in cavity 1 ($\alpha_1$) and 2 ($\alpha_2$). Together they form an order parameter $\alpha=\alpha_1+i\alpha_2=|\alpha | e^{i\phi}$ whose amplitude and phase directly map to the strength and the position of the density modulation. By detecting the photons leaking from the cavities we can continuously monitor the order parameter along both quadratures. 

The starting point of the experiment is an optically trapped, almost pure Bose-Einstein condensate of $2.02(6)\times 10^5$ $^{87}$Rb atoms that we expose to the transverse pump with wavelength $785.3\,\text{nm}$ and variable lattice depth.  We study the collective excitations of the quantum gas across the phase transition by probing it with cavity-enhanced Bragg spectroscopy, where probe photons are scattered off collective excitations into the transverse pump and vice versa. Our technique takes advantage of two key properties of optical cavities: enhancement of Bragg scattering and real-time access to the intracavity fields from leaking photons. The energy scale of the excitations is determined by the corresponding atomic recoil frequency $\omega_\text{rec}/2\pi= 3.7\,\text{kHz}$ for a transverse pump photon \cite{Nagy2008}. We prepare the system at a fixed transverse pump lattice depth and subsequently excite one of the two cavities with a probe field with time-varying detuning $\delta$ relative to the transverse pump frequency. In terms of the effective potential, this perturbs the order parameter along the quadrature of the probed cavity field (Fig.~\ref{fig:scheme}). We scan $\delta$ linearly in time from negative to positive detunings and record the intracavity photon numbers. Our technique can be regarded as a frequency-dependent extension of the method presented in \cite{Mottl2012}.

The probing situation qualitatively changes between the two phases, see Fig.~\ref{fig:spectrum}(A-C). In the normal phase, we probe the system on initially empty cavities and observe symmetric resonances at positive and negative $\delta$. These correspond to two-photon processes of probe and pump photons that involve the creation or annihilation of a density excitation in the system (Fig.~\ref{fig:spectrum}A). The circumstances are different in the ordered phase, in which the probe is applied on top of a finite order parameter and we observe a decay of the photon numbers in both cavities at a specific detuning $\delta$ (Fig.~\ref{fig:spectrum}B). We interpret this signal as a result of heating from an increased number of decaying excitations. The loss rate shows a symmetric resonance feature at positive and negative detunings. In addition, when probing weakly at detunings $\delta\ll\omega_\text{rec}$, a second resonance pair appears, see Fig.~\ref{fig:spectrum}C. Its visibility is highest when probing on an initially empty cavity. We therefore first enter the ordered phase in the presence of a symmetry breaking field along one quadrature and then ramp it down to zero before applying the probe field \cite{SI_excitations}. Over the entire coupling range covered by our measurements, we prepare the system in a weakly driven situation by adjusting the resulting mean intracavity photon numbers within a range of $\bar{n}_{1,2} = 0.06(1) - 3.4(2)$. 

\begin{figure}[t]
	\centering
		\includegraphics[width=\columnwidth]{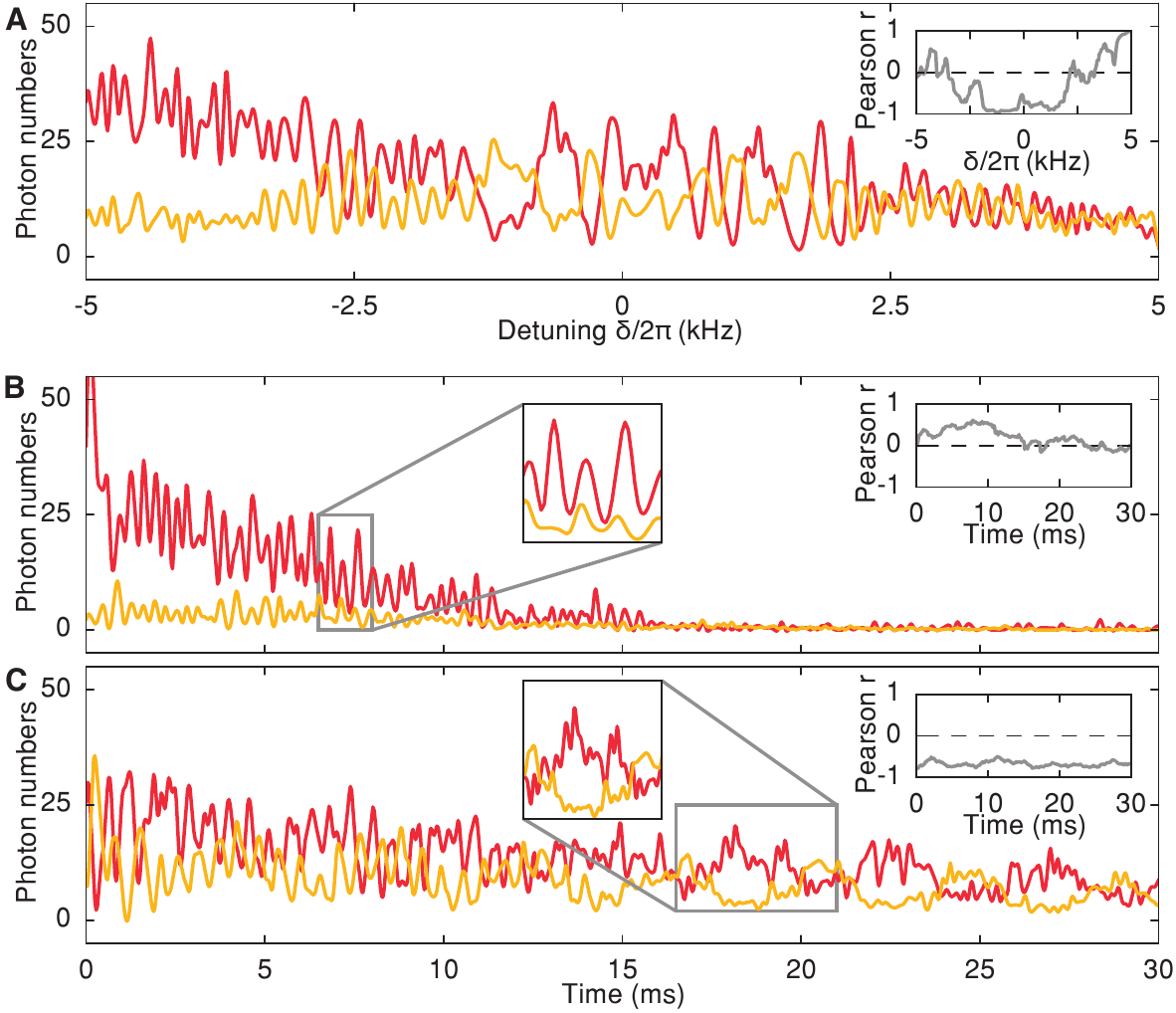}
	\caption{\textbf{Dynamics of Higgs and Goldstone excitations.} Response of the system at lattice depth $38.5(8)\,\hbar\omega_\text{rec}$ to a probe field on cavity 1. The insets show the Pearson correlation coefficient, deduced from a moving window of $2\,\text{ms}$ width of the mean photon numbers for cavity 1 (red) and 2 (yellow). \textbf{(A)} Ramp of the probe detuning $\delta$ with $0.5\,\text{kHz/ms}$ at constant probe field of $\bar{n}=0.03(1)$ intracavity photons in cavity 1. The correlation between the light fields is negative for $|\delta|/2\pi \lesssim2.5\,\text{kHz}$ and positive for higher $|\delta|$. \textbf{(B)} Evolution following a probe pulse of $1\,\text{ms}$ length with mean photon number $\bar{n}_1=2.4(2)$ in cavity 1 at $\delta/2\pi=2.5\,\text{kHz}$. We observe decaying correlations among the photon numbers, accompanied by a decreasing order parameter. \textbf{(C)} Evolution after the same probe pulse at $\delta/2\pi=0.5\,\text{kHz}$. Anticorrelations among the photon numbers decay without significiant loss of the order parameter. All data are binned in intervalls of $0.1\,\text{ms}$.}
	\label{fig:realtime}
\end{figure}

We record excitation spectra for different transverse pump lattice depths in the normal and the ordered phase. The signals can be obtained by probing either of the cavities. The resonance frequencies of the excitations are extracted from the spectra by fitting the data with a theoretical model \cite{SI_excitations}. The combined result is shown in Fig.~\ref{fig:spectrum}D. In the normal phase we observe decreasing resonance frequencies on approach to the critical point. When entering the ordered phase, two branches appear, one resonance remaining at frequencies small compared to $\omega_\text{rec}$, and a second one with rising frequencies. We find good agreement among the measurements for the two cavities over the entire covered range. The excitation frequencies can be well-described with a microscopic model \cite{SI_excitations}, which is related to previous theoretical work on spin systems with continuous symmetries \cite{Morrison2008a, Fan2014, Baksic2014}.

The separation of the excitation frequencies inside the ordered phase into a high and a low frequency branch suggests an interpretation in terms of a Higgs and a Goldstone mode. In order to carry out a direct test of the distinctive amplitude and phase character of the modes, we exploit the fact that the two cavity fields form the real and the imaginary part of the order parameter, thereby providing access to both quadratures. We prepare the system within the ordered phase at $38.5\,\hbar\omega_\text{rec}$ transverse pump lattice depth and apply a frequency-ramped probe field weak enough not to influence the lifetime of the system. The recorded evolution of the intracavity photon numbers is shown in Fig.~\ref{fig:realtime}A. We observe correlated signals for $|\delta|/2\pi\gtrsim 2.5\,\text{kHz}$ and anticorrelated signals at smaller probe detunings, consistent with the resonance frequencies shown in Fig.~\ref{fig:spectrum}. From their amplitude and phase character, we identify the excitations as Higgs and Goldstone modes. 

With regard to the atomic part of the excitations, these correspond to fluctuations in the strength and the position of the density modulation, as illustrated in Fig.~\ref{fig:scheme}C. The atomic coupling to delocalized cavity photons is equivalent to an effective atom--atom interaction of global range \cite{Ritsch2013, Asboth2004}. As a consequence, Higgs and Goldstone excitations are not inhibited, as it is the case in short-range interacting low-dimensional systems without long-range order \cite{Sachdev2011}. The global nature of the interaction furthermore results in a rigid crystal structure that inhibits the presence of excitations at non-zero wavenumbers, in contrast to theoretical studies on supersolid helium \cite{Liu1973}. The presence of a Higgs mode is ensured by the invariance of the system under an exchange of $\alpha_1$ and $\alpha_2$, which enforces a Lorentz-invariant time evolution \cite{SI_excitations} analogous to the particle-hole symmetry in e.\,g. superconductors and optical lattices at half-filling \cite{Pekker2015}. Finite-temperature effects are not expected to overdamp the Goldstone mode \cite{Piazza2013}.

Using the direct access to both quadratures of the order parameter we study the excitation dynamics induced by a strong probe pulse with constant detuning. Following a pulse of $1\,\text{ms}$ length at $\delta/2\pi=2.5\,\text{kHz}$, we observe correlated intracavity photon numbers signaling Higgs excitations, see Fig.~\ref{fig:realtime}B. The evolution of the light fields shows a damping of the Higgs modes over $\sim15\,\text{ms}$, accompanied by a decreasing order parameter. When applying a pulse at $\delta/2\pi=0.5\,\text{kHz}$ anticorrelated intracavity photon numbers are visible, showing the presence of Goldstone excitations, see Fig.~\ref{fig:realtime}C. The persisting low-frequency Goldstone mode is overlayed by a second fast-oscillating phase mode that decays within $\sim15\,\text{ms}$. 

\begin{figure}[t]
	\centering
		\includegraphics[width=\columnwidth]{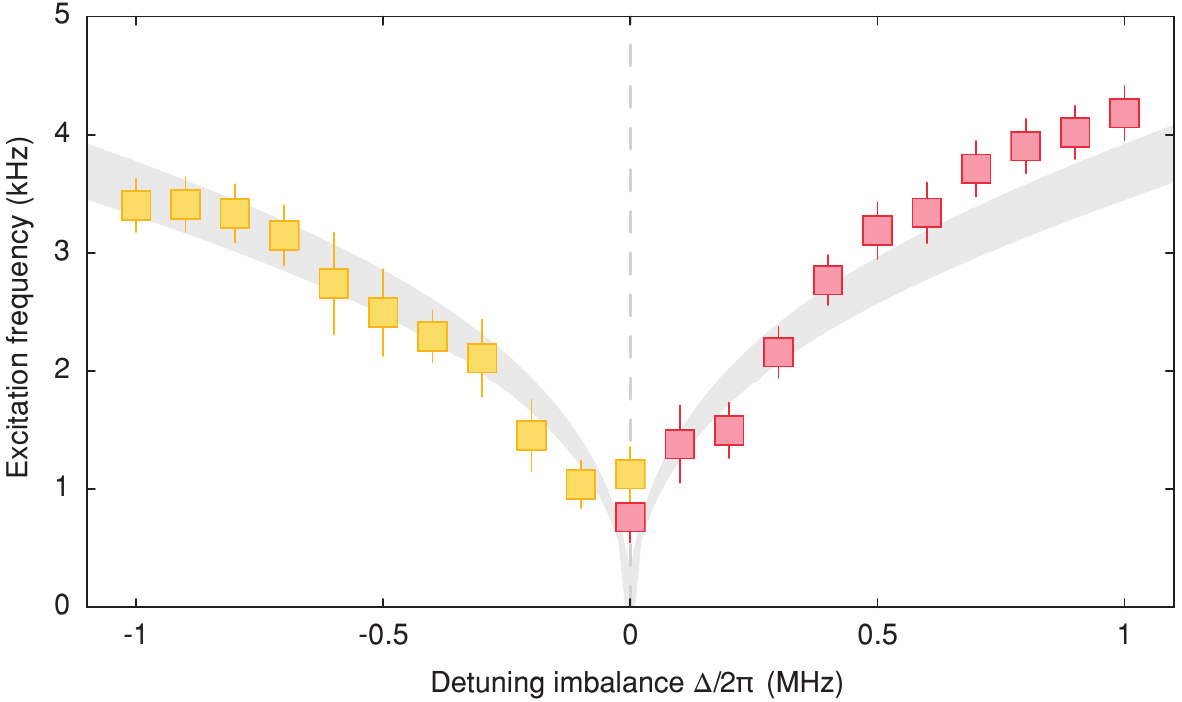}
	\caption{\textbf{Tunable mass of the Goldstone mode.} Resonance frequency as a function of the detuning imbalance $\Delta$. For positive (negative) $\Delta$, the mode couples to a probe field in cavity 1 (2) and its resonance frequency is shown in red (yellow). The resonance frequencies are derived from photon traces similar to Fig.~\ref{fig:spectrum}C, averaged over at least ten realizations. The gray-shaded area shows the theoretical prediction including experimental uncertainties \cite{SI_excitations}. The dashed line illustrates the situation of balanced coupling to both cavities. Errorbars combine fit errors and the uncertainty of the probe frequency.}
	\label{fig:mass}
\end{figure}

A hallmark of the Goldstone mode is its sensitivity to deviations from the continuous symmetry. Goldstone modes only show a vanishing excitation frequency for perfect symmetries in the absence of symmetry breaking fields or further interactions. An analogous behavior is known e.\,g.~in the context of chiral symmetry breaking, approximate symmetries, extra dimensions and the mass hierarchy problem \cite{Peskin1995, Arkani-Hamed1998}. The continuous symmetry that is broken in our system is the result of balanced coupling to two cavities, which each exhibit parity symmetry only. We can generate an adjustable symmetry breaking field along each quadrature of the order parameter individually by controlling the coupling to each cavity mode through its detuning from the transverse pump frequency \cite{SI_excitations}. For an imbalance $\Delta$ in the detunings, this results in an asymmetric effective potential with only two ground states on the axis of the more strongly coupled cavity field. The evolution of the resonance frequency of the Goldstone mode for various $\Delta$ around the balanced situation is shown in Fig.~\ref{fig:mass}. While it tends to zero for vanishing $\Delta$, we observe an increased resonance frequency for larger $| \Delta |$, approaching the soft mode associated to self-organization with a single cavity. The data are in agreement with our microscopic model. 

Thanks to the control over the effective potential landscape, our approach introduces a model system for studies on discrete and continuous symmetries. The unique real-time access to the system dynamics offers exciting prospects to examine the decay channels and coupling of Higgs and Goldstone modes \cite{Podolsky2011}. 

We thank J.~Larson, R.~Mottl, F.~Piazza and B.P.~Venkatesh for fruitful discussions, and are most grateful to W.~Zwerger for his insights and a careful reading of the manuscript. We acknowledge funding from Synthetic Quantum Many-Body Systems (European Research Council advanced grant) and the EU Collaborative Project TherMiQ (Grant Agreement 618074), and also SBFI support for Horizon2020 project QUIC and SNF support for NCCR QSIT and DACH project `Quantum Crystals of Matter and Light'.

\bibliographystyle{naturemag}
\bibliography{references}

\newpage


\setcounter{equation}{0}
\setcounter{figure}{0}
\renewcommand{\figurename}{Suppl. Fig.}

\section*{SUPPLEMENTARY INFORMATION}

\section*{Experimental details}
\subsection*{Setup and preparation of the Bose-Einstein condensate (BEC)}
We optically transport a cold thermal cloud of $^{87}$Rb atoms \cite{Leonard2014} along the $x$--axis into the cavity setup, where we apply forced optical evaporation to produce an almost pure BEC with $N=2.02(6)\times 10^5$ atoms. The atoms are retained in a dipole trap formed by two orthogonal laser beams at a wavelength of $1064\,\mathrm{nm}$ along the $x$-- and $y$--axes. The final trapping frequencies are $(\omega_x, \omega_y, \omega_z)=2\pi \times (66(1), 75(1), 133(5))\,\mathrm{Hz}$. The transverse pump is set to a wavelength $\lambda_\mathrm{p}=785.3\,\mathrm{nm}$, far red--detuned by $\Delta_a$ from the atomic $D_2$ line at $\lambda_\mathrm{D_2}=780.2\,\mathrm{nm}$. The cavities have single-atom vacuum Rabi frequencies $(g_1, g_2) = 2\pi\times (1.95(1), 1.77(1))\,\mathrm{MHz}$, and decay rates $(\kappa_1, \kappa_2) = 2\pi\times (147(4), 800(11))\,\mathrm{kHz}$. The cavities $i\in\{1,2\}$ are set to be resonant at $\omega_i$, detuned by $\Delta_i=\omega_\mathrm{p}-\omega_i$ from the pump frequency $\omega_\mathrm{p}$. Due to the mismatch in the cavity parameters, we achieve equal coupling to both cavities for $\Delta_1^{\mathrm{eq}}/2\pi=-3.2\,\mathrm{MHz}$ and $\Delta_2^{\mathrm{eq}}/2\pi=-2.9\,\mathrm{MHz}$ \cite{Leonard2017}.

\subsection*{Magnetic field to suppress spin transitions} We prepare the BEC in the $\ket{F, m_F}=\ket{1,-1}$ state with respect to the quantization axis along $\hat{z}$. The two cavities are birefringent with spacings of $(4.5, 4.8)\,\mathrm{MHz}$ between the horizontal and vertical eigenmodes. When fulfilling the resonance condition for a two-photon process involving pump and cavities, we can observe collective Raman transitions between different Zeeman sublevels accompanied by macroscopic occupation of the cavity modes. We suppress such spin changing scattering processes by applying a large magnetic offset field $B_z=34\,\mathrm{G}$, creating a Zeeman level splitting that is large compared to $\Delta_i$ and the birefringences.

\subsection*{Lattice calibrations} 
We calibrate the lattice depths of the transverse pump and each cavity field by performing Raman--Nath diffraction on the atomic cloud. The intracavity photon number can then be deduced from the calculated vacuum Rabi frequencies of the cavities. We extract efficiencies of $(5.0(2)\%,1.4(1)\%)$ for detecting an intracavity photon from cavity 1 and 2 with single--photon counting modules.

\subsection*{Excitation spectrum} 
For the measurements in the superfluid phase and for the Higgs mode, we fix the detunings at $(\Delta_1^{\mathrm{eq}}, \Delta_2^{\mathrm{eq}})$ and prepare the system at a given coupling strength by linearly increasing the transverse pump intensity within $50\,\mathrm{ms}$ to lattice depths up to $38.5(8)\,\hbar\omega_\mathrm{rec}$, with $\omega_\mathrm{rec}$ being the recoil frequency for a transverse pump photon. The measurements for the Goldstone mode were taken by first ramping up the transverse pump lattice within $30\,\mathrm{ms}$ at an imbalanced detuning of $(\Delta_1/2\pi, \Delta_2^{\mathrm{eq}}/2\pi) = (-4.0, -2.9)\,\mathrm{MHz}$ for the case of probing cavity 1 or $(\Delta_1^{\mathrm{eq}}/2\pi, \Delta_2/2\pi) = (-3.2, -3.7)\,\mathrm{MHz}$ for probing cavity 2, and then approaching the balanced situation at $(\Delta_1^{\mathrm{eq}}/2\pi, \Delta_2^{\mathrm{eq}}/2\pi)$ in a linear ramp of $20\,\text{ms}$ length. This effectively creates a symmetry breaking field during the preparation that sets the order parameter in the supersolid phase to have only one cavity populated. This way an empty cavity can be probed, which increases the signal quality. In some realizations the order parameter evolves during probing, and we post-select for zero order parameter in the probed cavity. The speed of the frequency ramp for the probe field is $0.2\,\mathrm{kHz/ms}$ for the Goldstone mode measurements and $1\,\mathrm{kHz/ms}$ for the other measurements. 

\subsection*{Excitation dynamics} 
We ramp up the transverse pump within $50\,\mathrm{ms}$ to a lattice depth of $38.5(8)\,\hbar\omega_\mathrm{rec}$ at $(\Delta_1^{\mathrm{eq}}, \Delta_2^{\mathrm{eq}})$ and perform the measurements shown in Fig.~3.

\subsection*{Tunable Goldstone mass}
A measurement for detuning imbalance $\Delta$ corresponds to the cavity detunings $(\Delta_1/2\pi, \Delta_2/2\pi) = (\Delta_1^{\mathrm{eq}} + \Delta/\sqrt2, \Delta_2^{\mathrm{eq}} - \Delta/\sqrt2)$. To prepare this measurement point, we first ramp up the transverse pump lattice within $30\,\mathrm{ms}$ at a far-imbalanced detuning of $(\Delta_1/2\pi, \Delta_2/2\pi) = (\Delta_1^{\mathrm{eq}} + 0.7\,\text{MHz}, \Delta_2^{\mathrm{eq}} - 0.7\,\text{MHz})$ for $\Delta>0$ and $(\Delta_1/2\pi, \Delta_2/2\pi) = (\Delta_1^{\mathrm{eq}} - 0.7\,\text{MHz}, \Delta_2^{\mathrm{eq}} + 0.7\,\text{MHz})$ for $\Delta<0$. We then approach the point for the measurement in a linear ramp. Afterwards, the probe frequency is ramped at a rate of $0.5\,\mathrm{kHz/ms}$ from $-5\,\mathrm{kHz}$ to $5\,\mathrm{kHz}$. 


\section*{Microscopic model}

\begin{figure}[t]
	\centering
		\includegraphics[width=\columnwidth]{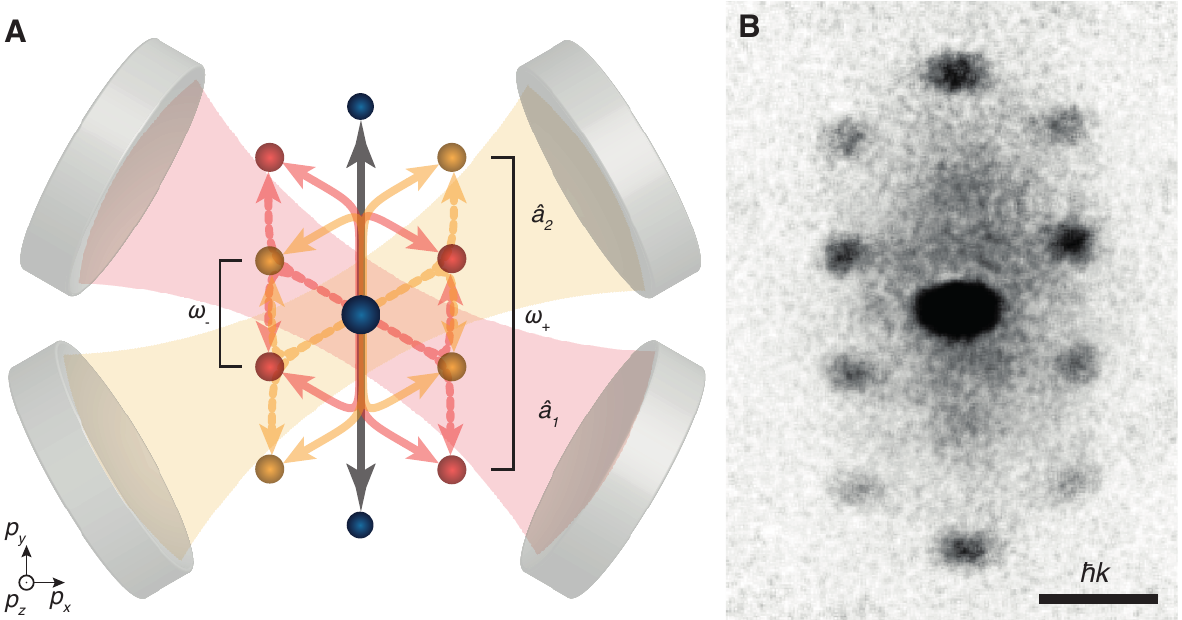}
	\caption{\textbf{Atomic momentum states. (A)} Coherent scattering processes of pump photons into cavity 1 (red) or cavity 2 (orange) and back give rise to atomic momentum states at energies $\hbar\omega_-=\hbar\omega_\mathrm{rec}$ and $\hbar\omega_+=3\hbar\omega_\mathrm{rec}$. Pump photons can also be scattered back into the pump (gray). The coordinate system is with respect to momentum space. \textbf{(B)} Absorption image of the atoms in the supersolid phase after $25\,\text{ms}$ ballistic expansion. All momentum states highlighted in \textbf{(A)} are populated.}
	\label{fig:momentum}
\end{figure}

\subsection*{Hamiltonian} Raman scattering between the pump and cavity fields via the atoms couples the atomic momentum state at $\ket{\textbf{k}}=\ket{\textbf{0}}$ to a superposition state of the higher momenta $\ket{\textbf{k}}=\ket{\pm \textbf{k}_\mathrm{p}\pm \textbf{k}_i}$, where $\textbf{k}_\mathrm{p}$ and $\textbf{k}_i$ denote the wave-vectors of the transverse pump and cavity $i$, respectively. These states fall into two groups with energy either $\hbar\omega_-=\hbar\omega_{\mathrm{rec}}$ or $\hbar\omega_+=3\hbar\omega_{\mathrm{rec}}$ (Fig.~\ref{fig:momentum}). As derived in the Methods section of \cite{Leonard2017}, our system is described by the effective Hamiltonian
\begin{equation}
\begin{aligned}
\hat{\mathcal{H}} =\sum_{i=1,2}\Bigl[&-\hbar\Delta_i\hat{a}^{\dagger}_i\hat{a}_i+\hbar\omega_{+}\hat{c}^{\dagger}_{i+}\hat{c}_{i+}+\hbar\omega_{-}\hat{c}^{\dagger}_{i-}\hat{c}_{i-}\\
&+\frac{\hbar\lambda_i}{\sqrt{N}}\Bigl(\hat{a}^{\dagger}_i+\hat{a}_i\Bigl)\Bigl(\hat{c}^{\dagger}_{i+}\hat{c}_{0}+\hat{c}^{\dagger}_{i-}\hat{c}_{0}+h.c.\Bigl)\Bigl].
\end{aligned}
\label{eq:Hamiltonian}
\end{equation}
$\hat{a}_i^\dagger$ ($\hat{a}_i$) are the creation (annihilation) operators for a photon in cavity $i$, $\hat{c}_{i\pm}^\dagger$ ($\hat{c}_{i\pm}$) and  $\hat{c}_0^\dagger$ ($\hat{c}_0$) create (annihilate) an atomic momentum excitation at energy $\hbar\omega_\pm$ associated with cavity $i$ and in the atomic ground state, respectively, $N$ is the atom number and $\lambda_i=\frac{\eta_i\sqrt{N}}{2\sqrt{2}}$ the Raman coupling which can be controlled via $\eta_i=-\frac{\Omega_\mathrm{p}g_i}{\Delta_a}$ with the transverse pump Rabi frequency $\Omega_p$. The dispersive shift $N g_i^2/(2\Delta_a)\ll\Delta_i$ is similar for both cavities and can be absorbed into $\Delta_i$. Other optomechanical terms can be discarded for our experimental parameters. As the vacuum Rabi coupling $g_i$ of both cavities are very similar, we use a single coupling $\lambda=\lambda_i$. 

\begin{figure*}[t]
	\centering
		\includegraphics[width=120mm]{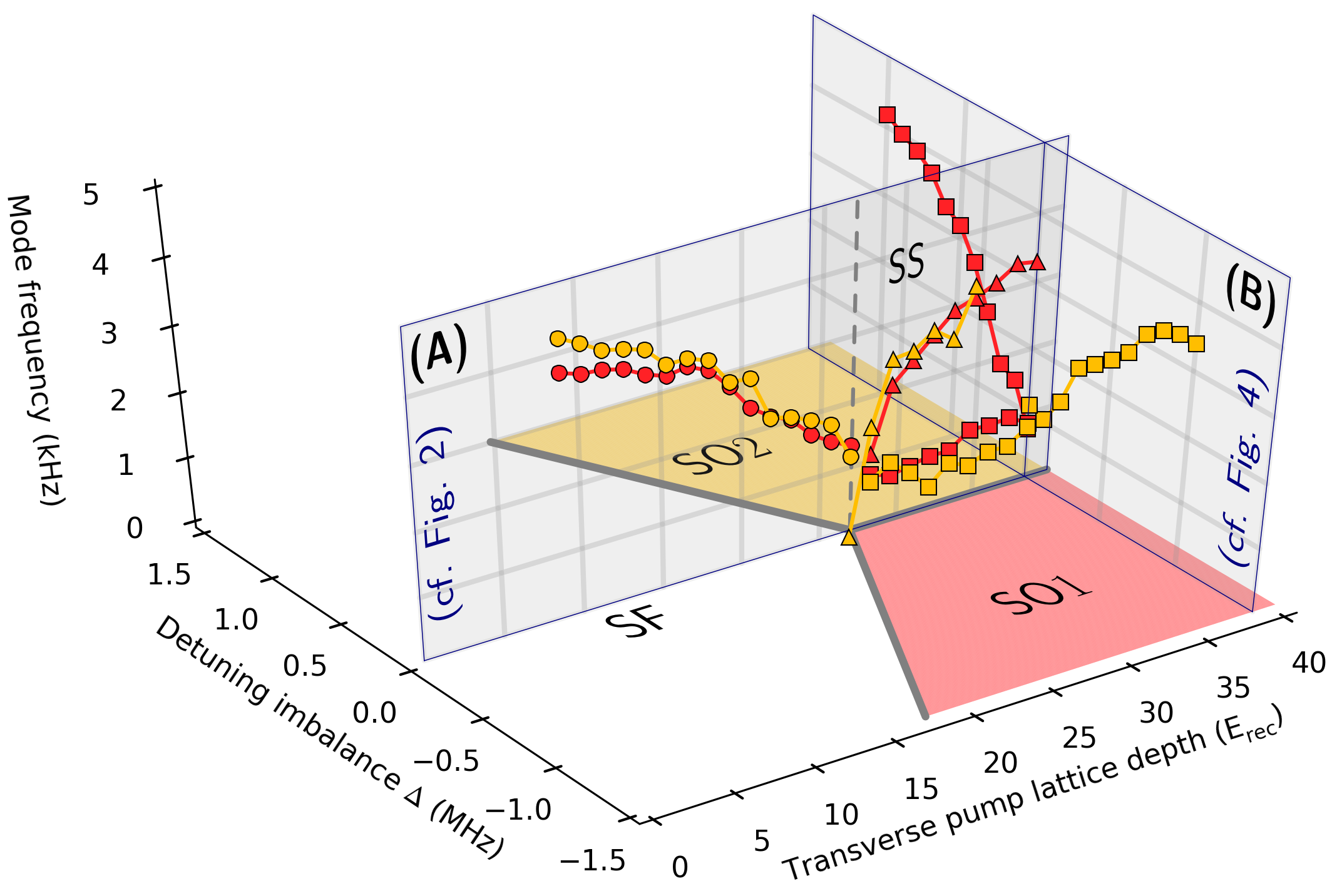}
	\caption{\textbf{Full phase diagram with excitation spectrum.} This figure shows the spectroscopic results of the paper in the context of the phase diagram of the system (bottom plane). For $\Delta=0$, an increasing transverse pump field leads to a phase transition from superfluid (SF) to supersolid (SS) (\textbf{A}--plane). The broken U(1) symmetry supports the massless Goldstone mode (lower branch) and massive Higgs mode (upper branch). Leaving the $\Delta=0$ line beyond critical coupling (\textbf{B}--plane), self-organisation is limited to only cavity 1 (SO1, $\Delta<0$) or cavity 2 (SO1, $\Delta>0$), accompanied by a crossover from continuous to discrete symmetry and a rise of the mass of the lower branch.}
	\label{fig:fancy3d}
\end{figure*}

\subsection*{Continuous symmetry} 
When coupling symmetrically to both cavities, i.\,e. $\Delta_C=\Delta_1=\Delta_2$, the Hamiltonian is symmetric under a simultaneous rotation in the basis of the light fields and the atomic fields. This can be captured by the generator
\begin{equation}
\hat{C}=-i\Bigl[\hat{a}_1^\dagger \hat{a}_2 - \hat{a}_2^\dagger \hat{a}_1 + \sum_{s=\pm}\Bigl(\hat{c}^{\dagger}_{1s} \hat{c}_{2s} - \hat{c}^{\dagger}_{2s} \hat{c}_{1s}\Bigl)\Bigl]. 
\end{equation}
It satisfies $[\hat{C},\hat{\mathcal{H}}]=0$, and, consequently, the Hamiltonian $\hat{\mathcal{H}}$ stays unchanged under the transformation $\hat{U}=e^{i\theta\hat{C}}$  for any $\theta\in\left[0,2\pi\right]$, i.\,e.\ $\hat{U} \hat{\mathcal H} \hat{U}^\dagger = \hat{\mathcal H}$. This U(1)--symmetry is broken at the phase transition. 

For arbitrary values of $\lambda$, $\Delta_1$ and $\Delta_2$, Eq.~\ref{eq:Hamiltonian} is instead $\mathbb{Z}_2$--symmetric under the operations $(a_i, c_{i+}, c_{i-})\rightarrow -(a_i, c_{i+}, c_{i-})$ for each individual cavity $i\in {1,2}$.

\subsection*{Phases} 
The cavity light field amplitudes $\alpha_i=\left<a_i\right>$ are order parameters describing the different phases of the system. A phase transition from the superfluid phase (SF, $\alpha_1=\alpha_2=0$) to a self-organized phase in cavity $i$ ($\mathrm{SO}1$ with $\alpha_1\neq0=\alpha_2$ and $\mathrm{SO}2$ with $\alpha_2\neq0=\alpha_1$) occurs when the coupling $\lambda$ crosses the critical coupling $\lambda^{\mathrm{cr}}_i = \sqrt{-\Delta_i\overline{\omega}/4}$, with $\overline{\omega}^{-1}=\omega_+^{-1}+\omega_-^{-1}$. This crossing is obtained by either changing $\lambda$ or $\Delta_i$. The supersolid phase ($\alpha_1, \alpha_2 \neq 0$) on the phase boundary between the $\mathrm{SO}1$ and the $\mathrm{SO}2$ phase is identified by the condition $\lambda^{\mathrm{cr}}_{1}=\lambda^{\mathrm{cr}}_{2}$, where the coupling to both cavities is symmetric. 

The phase boundaries as a function of the transverse pump lattice depth and the detuning imbalance $\Delta$ are shown on the bottom plane in Fig.~\ref{fig:fancy3d}. The spectra observed in Fig.~2 and 4 of the main text can be interpreted as the excitations for the supersolid phase and the SO1 and SO2 phases. 

\subsection*{Higgs and Goldstone modes in the supersolid phase} The Hamiltonian in Eq.~\ref{eq:Hamiltonian} can be solved numerically. We find that for cavity $i$, the state for the lowest eigenvalue has largest contribution from $\hat{c}_{i-}$, and the admixture of the $\hat{c}_{i+}$ mode is maximally $15\%$ in the explored parameter range, leading to a relative shift of the eigenfrequencies of $<10\%$ compared to including only the lowest mode. In order to derive an approximate analytic expression for the Higgs and Goldstone modes, we consider the following reduced Hamiltonian:
\begin{equation}
\begin{aligned}
\hat{\mathcal{H}} =\sum_{i=1,2}\Bigl[&-\hbar\Delta_i\hat{a}^{\dagger}_i\hat{a}_i+\hbar\omega_{-}\hat{c}^{\dagger}_{i-}\hat{c}_{i-}\\
&+\frac{\hbar\lambda}{\sqrt{N}}\Bigl(\hat{a}^{\dagger}_i+\hat{a}_i\Bigl)\Bigl(\hat{c}^{\dagger}_{i-}\hat{c}_{0}+\hat{c}_{0}^\dagger\hat{c}_{i-}\Bigl)\Bigl].
\end{aligned}
\label{eq:Hamiltonian_red}
\end{equation}
We describe the behavior of the atomic and the light modes by means of the Holstein-Primakoff transformations \cite{Hayn2011}:
\begin{equation}
\begin{aligned}
	&\hat{a}_i=\sqrt{N}\alpha_i+\delta \hat{a}_i \\
	&\hat{c}_{i-}=\sqrt{N}\psi_{i-}+\delta \hat{c}_{i-}\\
	&\hat{c}_0=\sqrt{N-\sum_{i=1,2}\hat{c}_{i-}^{\dagger}\hat{c}_{i-}}\\
\end{aligned}
\end{equation}
where $\delta \hat{a}_i$ ($\delta \hat{c}_{i-}$) describe the photonic (atomic) fluctuations of the system around its mean-field values $\alpha_i$ ($\psi_{i-}$). We expand the Hamiltonian in Eq.~\ref{eq:Hamiltonian_red} up to quadratic order in the excitations and use the quadratic part $\hat{h}^{(2)}=\hat{h}^{(2)}(\delta \hat{a}_i,\delta \hat{c}_{i-})$ to determine the excitation spectra of the system. In the normal phase ($\alpha_i=\psi_{i-}=0$), we obtain two orthogonal massive modes for the atomic excitations (see Fig.~1A)
\begin{equation}
	\omega_i = \omega_-\sqrt{1-\frac{\lambda^2}{\lambda_\text{cr}^2}}
	\label{eq:soft_mode}
\end{equation}
in the limit $\omega_{-}\ll |\Delta_C|$ with $\Delta_C=\Delta_1=\Delta_2$. Within the supersolid phase we have $\alpha_i,\psi_i\neq0$. Performing a rotation in the space of the excitations,
\begin{equation}
\begin{aligned}
	&\delta \hat{a}_1=\cos\theta\,\;\delta \hat{a}_{\mathrm H}+\sin \theta\,\;\delta \hat{a}_{\mathrm G}\\
	&\delta \hat{a}_2=-\sin\theta\,\;\delta \hat{a}_{\mathrm H}+\cos \theta\,\;\delta \hat{a}_{\mathrm G}\\
	&\delta \hat{c}_{1-}=\cos\theta\,\;\delta \hat{c}_{\mathrm H}+\sin \theta\,\;\delta \hat{c}_{\mathrm G}\\
	&\delta \hat{c}_{2-}=-\sin\theta\,\;\delta \hat{c}_{\mathrm H}+\cos \theta\,\;\delta \hat{c}_{\mathrm G},\\
\end{aligned}
\end{equation}
the quadratic part of the Hamiltonian separates into two contributions from the new modes $(\delta \hat{a}_{\mathrm G}, \delta \hat{a}_{\mathrm H},  \delta \hat{c}_{\mathrm G}, \delta \hat{c}_{\mathrm H})$,
\begin{equation}
\begin{aligned}
	&\hat{h}^{(2)}(\delta \hat{a}_{\mathrm G}, \delta \hat{a}_{\mathrm H},  \delta \hat{c}_{\mathrm G}, \delta \hat{c}_{\mathrm H})=\\ 
	&\hat{h}^{(2)}(\delta \hat{a}_{\mathrm G},  \delta \hat{c}_{\mathrm G})\,\;+\,\;\hat{h}^{(2)}(\delta \hat{a}_{\mathrm H}, \delta \hat{c}_{\mathrm H}).
\end{aligned}
\end{equation}
where
\begin{equation}
\begin{aligned}
	&\hat{h}^{(2)}(\delta \hat{a}_{\mathrm G},  \delta \hat{c}_{\mathrm G})= -\Delta_C\delta \hat{a}^{\dagger}_{\mathrm G}\delta\hat{a}_{\mathrm G}+\tilde{\omega}_-\delta\hat{c}^{\dagger}_{\mathrm G}\delta\hat{c}_{\mathrm G}\\
	&+\lambda\sqrt{\frac{1+\mu}{2}}\Bigl(\delta \hat{a}_{\mathrm G}^{\dagger}+\delta\hat{a}_{\mathrm G}\Bigl)\Bigl(\delta\hat{c}_{\mathrm G}^{\dagger}+\delta\hat{c}_{\mathrm G}\Bigl),
	\label{eq:microscopic_modes}
\end{aligned}
\end{equation}
and
\begin{equation}
\begin{aligned}
		\hat{h}^{(2)}(\delta \hat{a}_{\mathrm H}, \delta \hat{c}_{\mathrm H})= &-\Delta\delta \hat{a}^{\dagger}_H\delta\hat{a}_{\mathrm H}+\tilde{\omega}_-\delta\hat{c}^{\dagger}_H\delta\hat{c}_{\mathrm H}\\
	&+\frac{\omega_-(1-\mu)(3+\mu)}{4\mu(1+\mu)}\Bigl(\delta\hat{c}_{\mathrm H}^{\dagger}+\delta\hat{c}_{\mathrm H}\Bigl)^2\\
	&+\lambda\mu\sqrt{\frac{2}{1+\mu}}(\delta \hat{a}_{\mathrm H}^{\dagger}+\delta\hat{a}_{\mathrm H}\Bigl)\Bigl(\delta\hat{c}_{\mathrm H}^{\dagger}+\delta\hat{c}_{\mathrm H}\Bigl)
\end{aligned}
\end{equation}
with $\mu=(\lambda_\text{cr}/\lambda)^2$ and $\tilde{\omega}_-=\omega_-(1+\mu)/(2\mu)$.
$\hat{h}^{(2)}(\delta \hat{a}_G,  \delta \hat{c}_G)$ is of the form of the Hamiltonian in the normal phase  ($\alpha_i=\psi_i=0$)  where $(\tilde\omega_-^2-\Delta_C^2)^2+16\lambda^2\bigl[(1+\mu)/2\bigl]\Delta_C\tilde\omega_-=(\omega_-^2+\Delta_C^2)^2$. From this condition it directly follows that the excitation energy of this branch is zero \cite{Emary2003a}. On the other hand, $\hat{h}^{(2)}(\delta \hat{a}_H,  \delta \hat{c}_H)$ is of the form of the Hamiltonian that describes fluctuations around a superradiant phase with nonzero $\alpha_i,\psi_i$ and therefore a non--zero excitation energy. From this analysis we have shown that the excitation spectra in the supersolid phase separate into a gapped (Higgs) branch and a gapless (Goldstone) branch.

\begin{figure}[t]
	\centering
		\includegraphics[width=\columnwidth]{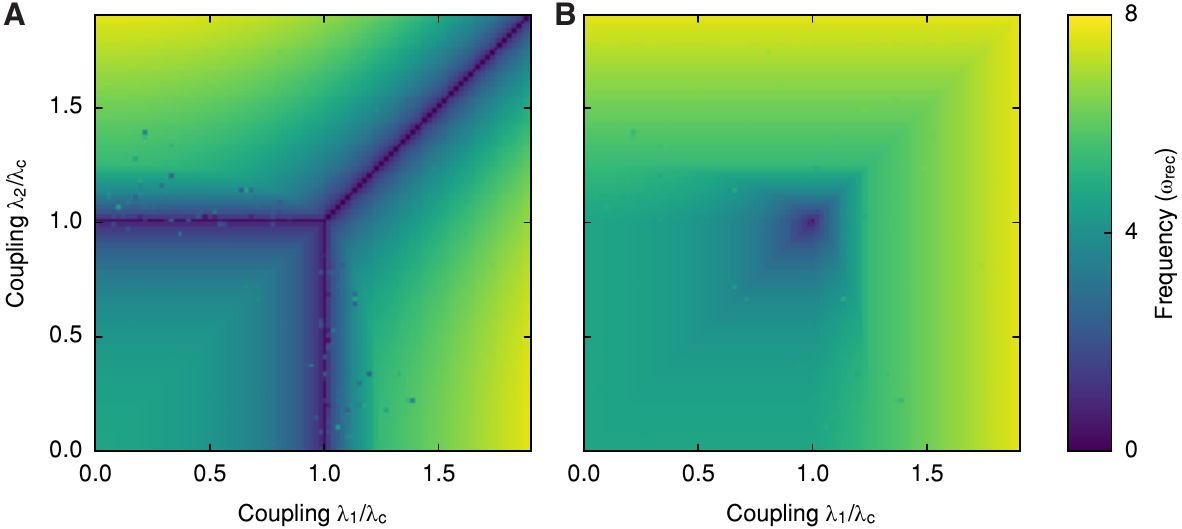}
	\caption{\textbf{Numerical results} Mean-field results for the lowest \textbf{(A)} and the second lowest \textbf{(B)} eigenmode. The couplings to each cavity are normalized by the critical coupling $\lambda_\text{cr}=\sqrt{-\Delta_C\bar{\omega}/4}$. In Fig.~2, we probe the system on the diagonal for $\lambda_1=\lambda_2$. The calculation includes all terms of the Hamiltonian in Eq.~\ref{eq:Hamiltonian} and additionally atomic contact interaction, the lattice potential and cavity decay. }
	\label{fig:mf_exc}
\end{figure}

\subsection*{Numerical mean--field solution} We can numerically obtain the eigenfrequencies of the full Hamiltonian in Eq.~\ref{eq:Hamiltonian} (including $\hat{c}_{i+}$) by expanding each operator around its expectation value and diagonalizing the Hessian matrix of the resulting mean--field Hamiltonian. The result for the lowest two excitations is shown in Fig.~\ref{fig:mf_exc}. Cavity decay, atom--atom contact interactions and the transverse pump potential are taken into account. The theory lines in Fig.~2 and 4 are obtained from this calculation, where uncertainties in the experimental parameters lead to the shaded gray area. We consider a 20\% systematic error in the atom numbers, fluctuations of the cavity resonance and the transverse pump laser frequency of 30\,kHz each, and the aforementioned uncertainty in the trapping frequency measurements.


\section*{Probing the excitation spectrum}
\subsection*{Time evolution of the excitations}
The Hamiltonian in Eq.~\ref{eq:Hamiltonian_red} separates into two parts that each describe one atomic mode coupled to a light mode. We can adiabatically eliminate the light fields and obtain the following effective Hamiltonian for each mode $\hat{c}_M$:
\begin{equation}
	\mathcal{\hat{H}}_\text{exc} = \sum_{M\in A,B} \hbar\omega_- \hat{c}_M^\dagger\hat{c}_M + \frac{\hbar \lambda^2}{N\Delta}\left( \hat{c}_M^\dagger\hat{c}_0 + \hat{c}_M\hat{c}_0^\dagger \right)^2.
\end{equation}
The operators $\hat{c}_M$ coincide with $\hat{c}_{i-}$ in the normal phase and are rotated in the ordered phase according to the outcome $\theta$ of the broken symmetry. In this section we explicitly include the decay rates $\kappa_i$ of the cavity fields into the calculation. A probe field on cavity $i \in \{1,2\}$ can be captured by
\begin{equation}
	\mathcal{\hat{H}}_\text{pr} = \hbar\xi(t)\sqrt{N}\left(\hat{c}_i^\dagger+\hat{c}_i \right) \cos(\delta t + \phi).
\end{equation}
Here, $\xi(t)=2\eta n_\text{pr}(t)$ is the probe field amplitude with mean intracavity photon number $n_\text{pr}(t)=\frac{\eta_\text{pr}^2}{\Delta_C^2+\kappa_i^2}$ for a cavity decay rate $\kappa_i$. The operator $\hat{c}_i^\dagger$ ($\hat{c}_i$) creates (annihilates) an atom in the excited state for cavity $i$. It can be decomposed in the excitation basis $\{\hat{c}_A, \hat{c}_B\}$. Similarly to \cite{Mottl2012}, this results in the time-dependent population of the excited state
\begin{equation}
	\langle\hat{c}_-^\dagger\hat{c}_-\rangle(t) = 4\eta^2n_\text{pr}N\xi\left[ \left(\frac{\omega_-}{\omega_\text{s}}\right)^2\Im(\mathcal{Y}(t))^2 + \Re(\mathcal{Y}(t)^2) \right]
	\label{eq:fit_function_atoms}
\end{equation}
with mode frequency $\omega_\text{s}$, $\hat{c}_-=\hat{c}_{1-} + i \hat{c}_{2-}$, and
\begin{equation}
	\mathcal{Y}(t)=e^{(i\omega_\text{s}-\gamma)t}\int^{t}_0{dt'}e^{-(i\omega_\text{s}-\gamma)t'}\cos(\delta t'+\varphi).
\end{equation}
The damping rate $\gamma$ is phenomenologically introduced to account for atomic decay. The corresponding photon field is
\begin{equation}
\begin{split}
	n_\text{ph}(t) = \bigg|\alpha - \frac{4\eta^2\sqrt{n_\text{pr}}N}{\Delta_i + i \kappa} \left(\frac{\omega_-}{\omega_\text{s}}\right)\Im(\mathcal{Y}(t)) + \sqrt{n_\text{pr}(t)}e^{-i(\delta t + \varphi)}\bigg|^2.
	\label{eq:fit_function}
\end{split}
\end{equation}
Since the relative phase $\varphi$ varies between realizations of the experiment, we perform an average $\langle n_\text{ph}(t)\rangle_{\varphi}$ over $\varphi\in\left[0,2\pi\right]$ and use the result as fit function for the response to a probe field with frequency $\delta$  relative to the transverse pump. We sweep $\delta=\delta(t)=\delta_0 + \delta't$ over time with rate $\delta'$. 

\begin{figure}[t!]
	\centering
		\includegraphics[width=\columnwidth]{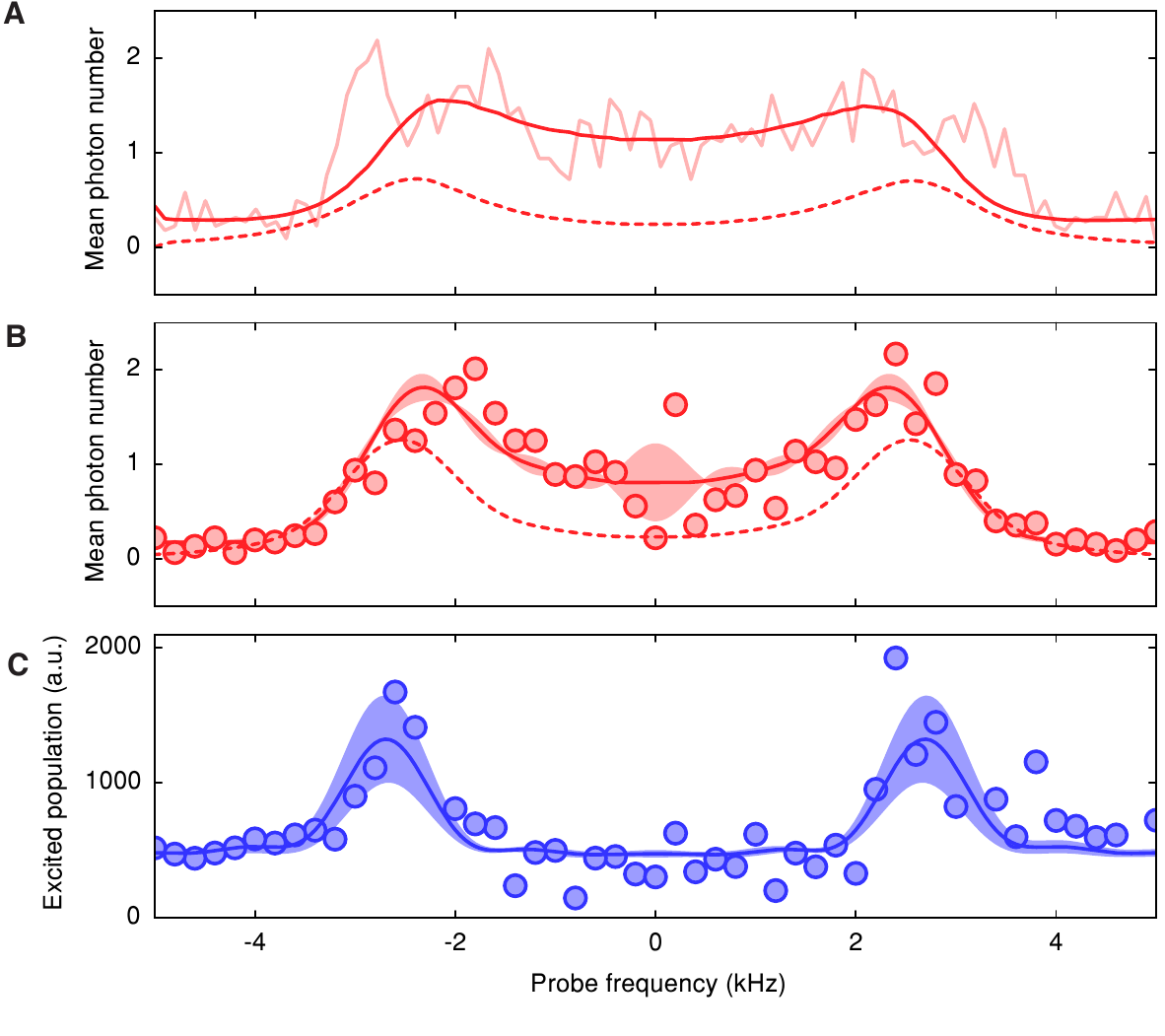}
	\caption{ \textbf{Comparison of probe methods.} We prepare the quantum gas in the normal phase at $\Delta_1=-2.0\,\text{MHz}$ and at a transverse pump lattice depth of $12.8(3)\,\hbar\omega_\text{rec}$. \textbf{(A)} Frequency ramp at a rate of $0.5\,\text{kHz/ms}$ with mean photon number $n=0.03(1)$ in cavity 1. We extract a resonance frequency of $2.97(10)\,\mathrm{kHz}$ from a fit with Eq.~\ref{eq:fit_function} with time-dependent detuning $\delta(t)$. \textbf{(B)} Each data point corresponds to the mean photon number during a probe pulse of $1\,\text{ms}$ length at constant detuning $\delta$ with mean photon number $n=0.07(1)$ in cavity 1. The solid line shows the fit with Eq.~\ref{eq:fit_function}, resulting in a resonance frequency of $2.92(5)\,\mathrm{kHz}$. \textbf{(C)} We extract the population $\langle \hat{c}_-^\dagger \hat{c}_-\rangle$ in the excited atomic state from pictures taken after $25\,\text{ms}$ ballistic expansion of the atoms after releasing them from the trap. The pictures are taken together with the data in \textbf{(B)}. The solid line is a fit of Eq.~\ref{eq:fit_function_atoms} to the data, resulting in a resonance frequency of $2.7(7)\,\mathrm{kHz}$. The dashed lines show the deduced photon number from Bragg scattering off the created excitations. Shaded areas denote the standard deviation of the fit function from the average over the phase $\varphi$. }
	\label{fig:probe_methods}
\end{figure}

Eq.~\ref{eq:microscopic_modes} implies that the time evolution for probing the Goldstone mode on a previously empty cavity is equivalent to probing the system in the normal phase. We therefore use $\langle n_\text{ph}(t)\rangle_{\varphi}$ as fit function to extract the resonance freqencies $\omega_s$ for the measurements in the normal phase and for the Goldstone mode in Fig.~2, as well as for Fig.~4. The fit parameters are $\omega_\text{s}$, $\sqrt{n_\text{pr}}$, $\eta$ and $\gamma$. For the Higgs mode measurements in Fig.~2, we use a Gaussian ansatz with free frequency, width, amplitude and offset.

\subsection*{Comparison of different probe methods} Instead of performing a frequency sweep, the excitation frequency can also be measured by perfoming pulses at constant frequency \cite{Mottl2012}. The method and the results are shown in Fig.~\ref{fig:probe_methods}. During a probe pulse, the intracavity photon number evolves over time and can be deduced by recording the photons leaking from the cavity. The atomic population in the excited momentum states $\hat{c}_{-}$ similarly evolves and its value at the end of the probe pulse can be extracted from absorption images of the cloud after ballistic expansion. We model the population $\langle\hat{c}_-^\dagger\hat{c}_-\rangle_\varphi$ and extract the resonance frequency from a fit to the data. The results from all three measurement methods agree within the experimental uncertainties.  


\section*{Effective action}
\subsection*{Effective potential} In order to derive an effective potential for our system, we start by inserting the mean-field ansatz $\langle \hat{a}_i \rangle=\sqrt{N}\alpha_i$, $\langle \hat{c}_{i-} \rangle=\sqrt{N}\psi_i$ and $\langle \hat{c}_0 \rangle=\sqrt{N\left( 1-\psi_1^2-\psi_2^2 \right) }=\sqrt{N}\psi_0$ into the reduced Hamiltonian in Eq.~\ref{eq:Hamiltonian_red}, with $\alpha_i$, $\psi_i\in \mathbb{R}$. This results in the effective potential
\begin{equation}
	V(\alpha_i, \psi_i) =\sum_{i=1,2}\Bigl[-\hbar\Delta_i\alpha_i^2+\hbar\omega_-\psi_i^2+4\hbar\lambda\alpha_i\psi_i\psi_0\Bigl]\text{.}
\end{equation}
As $\Delta_i\gg\omega_-$, the photon fields reach their steady state quasi-instantaneously with respect to the atomic fields. We can hence adiabatically eliminate the light fields enforcing the condition $\frac{\partial V}{\partial \alpha_i} = 0$ and obtain
\begin{equation}
	V(\psi_1,\psi_2) = \hbar \omega_- (\psi_1^2+\psi_2^2) + 4\hbar\lambda^2\psi_0^2 \left( \frac{\psi_1^2}{\Delta_1} + \frac{\psi_2^2}{\Delta_2} \right).
\end{equation}
For the U(1)--symmetric case with $\Delta_C=\Delta_1=\Delta_2$, this simplifies to
\begin{equation}
	V(\psi) = \hbar \omega_- \left[|\psi|^2\left( 1 - \frac{\lambda^2}{\lambda_\text{cr}^2}\right) + |\psi|^4 \frac{\lambda^2}{\lambda_\text{cr}^2} \right]\text{,}
	\label{eq:effective}
\end{equation}
where $\psi = \psi_1 + i \psi_2$  is a complex atomic order parameter and $\lambda_\text{cr} = \sqrt{-\Delta_C \omega_-/4}$. For $\lambda<\lambda_\text{cr}$, the potential has a 'bowl' shape as displayed in Fig.~1A. For $\lambda>\lambda_\text{cr}$, it acquires a 'sombrero' shape with a circular manifold of minima at 
\begin{equation}
	\left|\psi_0\right| = \sqrt{\frac{1}{2}\left(1 - \frac{\lambda_\text{cr}^2}{\lambda^2}\right)}
	\label{eq:mf_solution}
\end{equation}
This potential can be probed along $\psi_1$ and $\psi_2$ independently thanks to the expression $\alpha_i=2\lambda\psi_i\psi_0/\Delta_C$, obtained from the condition $\frac{\partial V}{\partial \alpha_i} = 0$. The combined cavity field $\alpha = \alpha_1 + i \alpha_2$ therefore constitutes an equivalent order parameter. 

\subsection*{Lorentz--invariant effective action} In order to describe the elementary excitations, we need to add dynamic terms to the effective potential. The leading orders of the time evolution of the effective action in Landau-Ginzburg theory are
\begin{equation}
	\mathcal{S}_{\mathrm{dynamic}} = K_1 (\psi^*\partial_t \psi - \psi \partial_t \psi^*) + K_2 \partial_t \psi^* \partial_t \psi\text{,}
\end{equation}
with two scalar constants $K_1$ and $K_2$. With $\psi=\psi_1+i\psi_2$ we get
\begin{equation}
	\mathcal{S}_{\mathrm{dynamic}} = 2iK_1 (\psi_1\partial_t\psi_2 - \psi_2\partial_t\psi_1) + K_2 \left[(\partial_t\psi_1)^2 + (\partial_t\psi_2)^2\right]\text{.}
\end{equation}
For our system, the Hamiltonian is invariant under the exchange of the two cavities, i.\,e.~$\alpha_1$ and $\alpha_2$. Accordingly, the same invariance holds for the momentum modes $\psi_1$ and $\psi_2$. If we require the effective action to fulfill the same condition, it follows that $K_1=0$. This argument is analogous to the situation e.\,g.~in superconductors or at the superfluid/Mott--insulator phase transition in optical lattices with half filling, where a particle--hole symmetry enforces the effective action to be Lorentz--invariant\cite{Pekker2015}.

\subsection*{Higgs and Goldstone modes}
The effective action allows to directly deduce the Higgs and the Goldstone modes. In the zero-momentum limit, the frequency of the Higgs mode follows the condition \cite{Pekker2015}:
\begin{equation}
	\omega^2 = \frac{1}{K_2}\frac{\partial^2 V}{\partial\psi^2}
	\label{eq:exc_freqs}
\end{equation}
From the analytic expression for the normal phase in Eq.~\ref{eq:soft_mode}, we can conclude $K_2=2\hbar/\omega_-$. Assuming $K_2$ constant throughout the covered range, Eq.~\ref{eq:exc_freqs} allows to obtain the frequencies $\omega_H$ for the Higgs and $\omega_G$ for the Goldstone mode in the ordered phase. With the Laplace operator in polar coordinates
\begin{equation}
	\frac{\partial^2 V}{\partial \psi^2} = \frac{\partial^2 V}{\partial\left|\psi\right|^2} + \frac{1}{\left|\psi\right|}\frac{\partial V}{\partial\left|\psi\right|} + \frac{1}{\left|\psi\right|^2}\frac{\partial V^2}{\partial \phi^2}
\end{equation}
and using Eq.~\ref{eq:mf_solution}, we obtain:
\begin{align}
	\omega_H &= \sqrt{2}\omega_-\sqrt{\frac{\lambda^2}{\lambda_\text{cr}^2}-1} \\
	\omega_G &=0
\end{align}

\end{document}